\documentclass[journal]{IEEEtran}
\usepackage{amsmath,amsfonts}
\usepackage{array}
\usepackage{textcomp}
\usepackage{stfloats}
\usepackage{url}
\usepackage{verbatim}
\usepackage{graphicx}
\usepackage{cite}

\usepackage{algorithm}
\usepackage{algpseudocode}
\usepackage{multirow}

\usepackage{pifont}
\usepackage{threeparttable}
\usepackage{booktabs}
\usepackage{ragged2e}
\usepackage{makecell}
\usepackage{subcaption}
\usepackage[justification=centering]{caption}
\usepackage{xspace}

\usepackage{soul,color}

\newcommand{\name}{\texttt{FreeTalk}\xspace}

\begin{document}

\title{\texttt{FreeTalk}:A plug-and-play and black-box defense against
speech synthesis attacks}

\author{Yuwen Pu, Zhou Feng, Chunyi Zhou, Jiahao Chen, Chunqiang Hu, Haibo Hu, Shouling Ji
\thanks{Yuwen Pu, Chunqiang Hu and Haibo Hu are with the School of Big Data \& Software Engineering, Chongqing University, Chongqing, 400030, China. E-mail: \{yw.pu, chu, haibo.hu\}@cqu.edu.cn.}
\thanks{Zhou Feng, Chunyi Zhou, Jiahao Chen and Shouling Ji are with the College of Computer Science and Technology at Zhejiang University, Hangzhou, Zhejiang, 310027, China. E-mail: \{zhou.feng, zhouchunyi,  xaddwell, sji\}@zju.edu.cn}
\thanks{Yuwen Pu and Zhou Feng are the co-first authors.}}


\maketitle


\begin{abstract}
Recently, speech assistant and speech verification have been used in many fields (such as smart home, finance, speech translation, and audio navigation), which brings much benefit and convenience for us. However, when we enjoy these speech applications, our speech may
be collected by attackers for speech synthesis. For example, an attacker generates some inappropriate political opinions with the characteristic of the victim's voice by obtaining a piece of the
victim’s speech, which will greatly influence the victim's reputation. Specifically, with the appearance of some zero-shot voice conversion methods, the cost of speech synthesis attacks has been further reduced, which also brings greater challenges to user voice security and privacy. Some researchers have proposed the corresponding privacy-preserving methods. However, the existing approaches have some on-negligible drawbacks: low transferability and robustness, high computational overhead. These deficiencies seriously limit the existing method deployed in practical scenarios. Therefore, in this paper, we proposed a lightweight, robust, plug-and-play privacy preservation method against speech synthesis attacks in a black-box setting. Our method generates and adds a frequency-domain perturbation to the original speech to achieve privacy protection and high speech quality. Then, we present data augmentation strategy and noise smoothing mechanism to improve the robustness of the proposed method. Besides, to reduce the user's defense overhead, we also proposed a novel identity-wise protection mechanism. It can generate a universal perturbation for one speaker and support privacy preservation for speech of any length. Finally, we conduct extensive experiments on 5 speech synthesis models, 5 speech verification models, 1 speech recognition model, and 2 datasets. The experimental results demonstrate that our method has satisfying privacy-preserving performance, high speech quality, and utility. 
\end{abstract}


\section{Introduction}

\IEEEPARstart {R}ecent advances in speech synthesis have led to powerful techniques like Voice Conversion (VC) \cite{liu2021any, sheng2023face} and Text-to-Speech (TTS) \cite{yu2023antifake}.  VC transforms the speaker's voice characteristics while preserving linguistic content\cite{deng2023pmvc}, whereas TTS generates natural speech from text input using a target speaker's voice. These technologies are widely applied in voice assistants, language translation, and assistive communication tools \cite{james2021empathetic, jia1904direct, mertl2018quality, nakamura2009overcoming}.

While speech synthesis technologies provide considerable convenience in human-computer interaction, they simultaneously raise critical concerns regarding security vulnerabilities and privacy infringements. A growing concern involves malicious actors exploiting these systems to fabricate artificial voices for fraudulent or infringing activities \cite{chen2024adversarial, chen2024proactive}. Notably, emerging zero-shot VC techniques \cite{shen2023naturalspeech, huang2022generspeech, kang2022end} enable attackers to synthesize highly realistic voice replicas using merely seconds of victim audio samples obtained from social media platforms. Such technological advancements have led to increasingly sophisticated attack vectors, as exemplified by a case where synthesized CEO voice audio defrauded victims of over $\$243,000$ through telephone deception \cite{stupp2019fraudsters, liu2023protecting}.

\begin{figure}[h]
\centering
\includegraphics[width=1\linewidth]{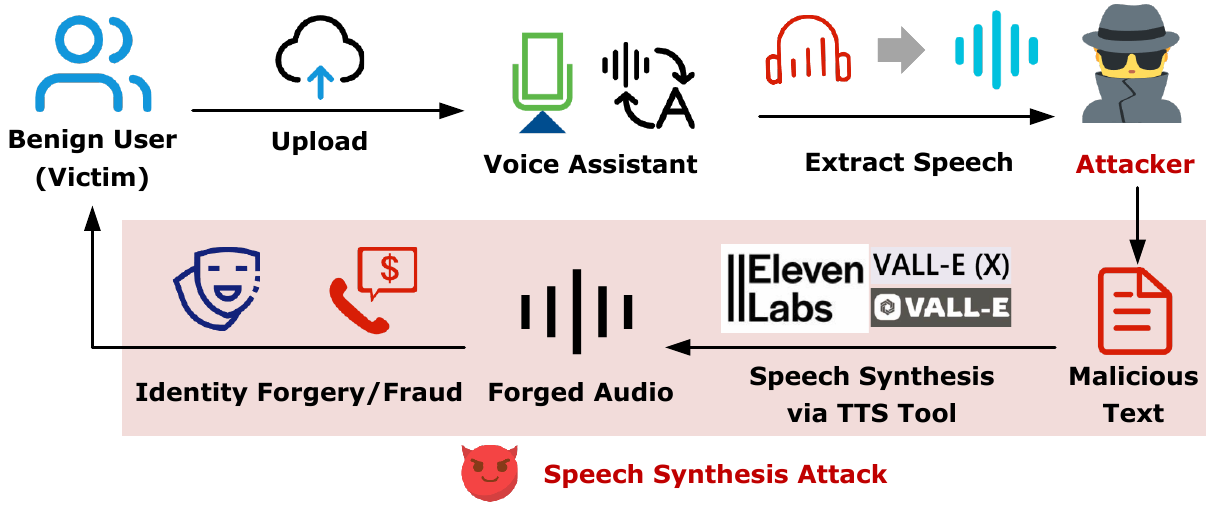}
\caption{The overview of speech synthesis attack.}
\label{fig:VC}
\end{figure}


To counter these emerging threats, current research has primarily focused on developing adversarial-based defense mechanisms \cite{liu2023protecting, yu2023antifake, chen2024adversarial, chen2024proactive}. They mainly protect user's voice from speech synthesis attacks by adding some adversarial perturbations to original audio. Although these defensive methods have achieved some success in combating voice spoofing, they often suffer from these \textit{significant drawbacks}. \textit{(i)} \textbf{Low Transferability and Strong Assumptions}. Current defense methods assume knowledge of the attack model, an unrealistic assumption given that users cannot know which specific model surveillants use in practice \cite{chen2024adversarial, yu2023antifake, liu2023protecting}. Consequently, these adversarial perturbations demonstrate limited cross-model transferability, although they may perform better on known attack models - an overly strong assumption in practical scenarios. \textit{(ii)} \textbf{Limited Generalizability}. Existing methods require unique computations for each speech segment \cite{yu2023antifake, liu2023protecting}, fundamentally restricting their generalizability across different usage scenarios. This segment-specific approach results in poor scalability across input durations, as the framework cannot effectively generalize its protective mechanisms to varying waveform lengths. The consequent computational inefficiency emerges as a direct manifestation of this core limitation, particularly evident in resource-constrained deployment environments where generalizable solutions are paramount.




Motivated by these gap, we introduce \name, which is a plug-and-play and black-box defense approach against speech synthesis attacks. Using \name, although the processed sample maintains perceptual similarity to the victim's voice for human listeners, synthesis using this sample by attacker generates speech that diverges from the victim's vocal characteristics. Consequently, the synthesized speech cannot deceive authentication services or the victim's acquaintances, while preserving the victim's ability to share speech normally on voice assistant platforms. \name can be seamlessly deployed on major voice assistant platforms as a plug-and-play solution, providing universal audio protection without compromising perceptual quality. This enables users to upload audio content \textbf{freely} and \textbf{securely} without concerns over privacy or auditory degradation.

Designing \name presents three main technical challenges:

\begin{enumerate}
    \item \textbf{Transferability.}  Considering the limited background knowledge of practical defenders, it is difficult for them to determine the type of the attack model, making it challenging for perturbation methods to achieve cross-model transferability. \textit{How can we generate perturbations with good transferability against unknown attack types in black-box scenarios?}
    \item \textbf{Robustness.} Current methods, while ensuring security by adding noise to audio, often significantly degrade usability (e.g., timbre). However, overly small perturbations provide insufficient protection. \textit{How can we strike an optimal trade-off, ensuring robust speech protection while maintaining high-precision speech recognition?}
    \item \textbf{Generalizability.} Current methods predominantly apply noise injection to individual speech segments, incurring substantial computational overhead - particularly when processing lengthy protected speech.  \textit{How can we design an efficient identity-wise perturbation mechanism that produces universal noise patterns applicable to all speech instances from a given user?}
\end{enumerate}

To overcome these challenges, we introduce the first plug-and-play and black-box defense against speech synthesis, namely \name. For challenge 1), \name equips with a novel sample-wise perturbation mechanism that applies frequency-domain perturbations to original speech signals, achieving improving transferability in the context of black-box and unknowing attack scenarios. For challenge 2), \name combines a novel data augmentation strategy with an advanced noise smoothing mechanism, specifically designed to eliminate high-frequency distortions while preserving the spectral integrity of protected speech signals.  For challenge 3), \name introduces a hierarchical patch optimization method combining identity-wise and frame-wise processing, enabling both speaker-universal perturbation generation and length-agnostic speech preservation. The specific contributions of this paper are outlined as below:


\begin{enumerate}
    \item We propose \name, the first black-box, plug-and-play speech protection framework that effectively defends against speech synthesis attacks. The framework could demonstrate strong transferability against unknown attack models.
    \item We design a novel data augmentation strategy to enhance robustness with a specialized noise smoothing mechanism. It effectively eliminates high-frequency distortions while preserving the spectral integrity of protected speech signals.
    
    \item We present a hierarchical patch optimization method that integrates identity-wise and frame-wise processing to enhance generalizability, enabling both speaker-universal perturbation generation and length-agnostic speech preservation.

    \item We evaluate our \name on 5 speech synthesis models, 5 speech verification models, 1 speech recognition model, and 2 datasets. Experimental results demonstrate that \name achieves strong privacy protection while maintaining high speech quality and utility.

\end{enumerate}

\section{Related Works}
In this section, we review the relevant works on speech synthesis methods and introduce the development of voice conversion defense methods.
\subsection{Speech Synthesis Attacks}
In speech synthesis attacks, the attacker aims to generate a fake speech with the target speaker's voice identity, speaking some specific words selected by the attacker. In this paper, we mainly focus on two popular speech synthesis attacks, including VC and TTS.
\subsubsection{Voice Conversion}
VC can be categorized into parallel VC and nonparallel VC based on whether they require a parallel corpus for training. In the early stages of VC development, many parallel VC approaches are proposed \cite{toda2007voice, turk2006robust, sun2015voice}. For example, based on the transformer architecture, Liu et al. \cite{liu2020voice} proposed a VC model that can generate high quality speech with high similarity to the target speaker. Helander et al. \cite{helander2010voice} proposed a VC approach by combining the partial least squares (PLS) with the Gaussian mixture model (GMM), which allows the use of multiple local linear mappings. However, in the real scenario, it is impractical to obtain a number of parallel training data. Therefore, many non-parallel VC methods are proposed\cite{hsu2016voice, wu2020one, nakashika2016non, saito2018non,chen2021again}. For example, Liu et al. \cite{liu2021any} designed a robust nonparallel and supporting any-to-any VC method. This method can achieve satisfactory performance in terms of naturalness and target speaker similarity. Kang et al. \cite{kang2022end} presented an end-to-end zero-shot voice conversion model that can disentangle content and timbre information and then efficiently synthesize the time domain audio. Nguyen et al. \cite{nguyen2022nvc} devised an end-to-end adversarial network that supported conducting VC directly on arbitrary length raw audio waveform and had a fine performance. Wang et al. \cite{wang2023lm} proposed a large-model-based zero-shot VC method that used a two-stage framework. \name can easily generate the fake speech with only 3 seconds of the target speaker's voice.

\subsubsection{Text-to-Speech}
 Similar to VC, TTS generates a target utterance with arbitrary texts and the timbre of the target speaker. Now, many advanced TTS methods are also proposed \cite{jia2018transfer, ren2019fastspeech, hu2019neural}. For example, Wang et al. \cite{wang2023neural} introduced a language model approach for TTS with audio codec codes as intermediate representations. This approach can synthesize high-quality personality speech with only a few seconds speech from the target speaker. Shen et al. \cite{shen2023naturalspeech} designed an efficient TTS system that uses a neural audio codec with continuous latent vectors to achieve natural and zero-shot TTS synthesis. Huang et al. \cite{huang2022generspeech} proposed a TTS model that disentangled the speech into style-agnostic and style-specific information. This model had a good performance with high audio quality and style similarity.
 
\subsection{Defending against Speech Synthesis Attacks}
To mitigate speech synthesis attacks, many defense methods have been proposed, including two directions: post-detection \cite{ahmed2020void, albadawy2019detecting, blue2022you,liu2023detecting} and preventing the attacker from generating synthetic speeches \cite{liu2023protecting, zhang2024hiddenspeaker, huang2021defending, feng2025enkidu}. Note that in this paper, we mainly focus on how to protect speakers' voice identity privacy by preventing the attacker to generate synthetic speeches. There are some relevant privacy-preserving methods that mainly employ the idea of adversarial example generation. For example, Huang et al. \cite{huang2021defending} prevented the speaker's voice characteristics from leaking by introducing some human imperceptible noise into the utterances of a speaker. Chen et al. \cite{chen2024adversarial} proposed a speaker protection method that perturbs speech signals by minimally altering the original speech while preventing the attacker from accurately generating the voice of the target speaker. Note that both of the two above methods perform well only in the white-box scenario. Some methods are also proposed for both white-box and black-box scenarios. For example, Liu et al. \cite{liu2023protecting} proposed two defense methods that can significantly degrade the performance of speech synthesis attacks while maintaining the sound of the speaker's voice, but the two defense methods have to obtain the output of the attack model given an input. Yu et al. \cite{yu2023antifake} designed a defense mechanism that relies on the transferability of adversarial examples to prevent unauthorized speech synthesis, but assumed that the attack model must share similarities with other robust encoders for improved efficacy. Li et al. \cite{li2023voice} proposed a voice guard defense method that added the adversarial perturbation to the time domain. 
Dong et al. \cite{dong2024active} presented a Generative Adversarial Network (GAN) framework to defend against malicious speech synthesis, which had a fine defense effectiveness and generation time. However, all the above methods perform well in the white-box setting and have poor performance in black-box. \cite{zhang2024hiddenspeaker} and \cite{ge2022wavefuzz} also proposed to perturb the speaker's voice to make them unlearnable for other deep-learning-based audio models. Note that all existing methods must generate a specific perturbation for each speech of the same speaker, which significantly increases computational overhead. 

\section{Threat Model}

\subsection{Attacker Motivation and Assumptions}
\noindent\textbf{Attack Motivation}: The attacker intends to leverage the victim's voice samples, which are uploaded to voice assistant tools (such as speech-to-text converters, translators, etc.), to synthesize speech using TTS tools. By imitating the victim's vocal characteristics, the attacker can generate speech with manipulated content, thereby engaging in malicious activities such as fraudulent blackmail or reputational damage, ultimately causing severe harm to the victim.

\textit{(1) Financial Motivation.} Attackers primarily seek monetary gain through voice spoofing, including financial fraud (e.g., fraudulent transfers), ransom extortion, and bypassing voice authentication systems in banking or corporate environments. Commercial espionage may also occur through executive voice impersonation, as exemplified by a case where synthesized CEO voice audio defrauded victims of over $\$243,000$ through telephone deception \cite{stupp2019fraudsters, liu2023protecting}.

\textit{(2) Sociopolitical Motivation.} This involves manipulating public perception by fabricating damaging voice content to harm reputations of individuals or organizations. Politically, such attacks can spread disinformation by impersonating public figures to influence elections or international relations. During the Russia-Ukraine conflict, a fabricated audio of President Zelensky ``ordering soldiers to surrender'' circulated on social media to undermine Ukrainian troops' morale \cite{Russia}.

\textit{(3) Personal Motivation.} Attackers pursue personal objectives like revenge or harassment through targeted voice manipulation. This includes creating threatening messages or fabricating emotional distress scenarios to exploit personal relationships.

\noindent\textbf{Attacker's Assumptions}: We assume that the attacker can obtain some victim users' voice snippets, such as by eavesdropping from voice assistant tools, which is relatively easy to achieve because numerous existing studies have demonstrated privacy leakage vulnerabilities in current voice-based systems\cite{liu2023privacy,cheng2022personal}. Alternatively, the attacker may directly collect voice samples if they have a prior acquaintance with the victim. The attacker also has enough computational resource to run the SOTA speech synthesis methods (e.g., zero-shot speech synthesis) to generate fake speech of the target speaker.


\subsection{System Goals and Defender Assumptions}

\noindent\textbf{Defender's Goal}: 
The defender employs voice assistant tools or voice translator for services. To defend against potential threats, \name applies security processing to the uploaded speech. This ensures that automatic speech recognition and other voice-based services function normally while preventing attackers from synthesizing high-value voice samples that mimic the victim’s timbre using voice generation tools.

\noindent\textbf{Defender's Assumptions}: 
The defender possesses no prior knowledge of the attacker's methodologies or models. The defender’s capability is limited to processing the shared speech using the proposed \name prior to public release.

\begin{figure*}[h]
\centering
\includegraphics[width=0.8\linewidth]{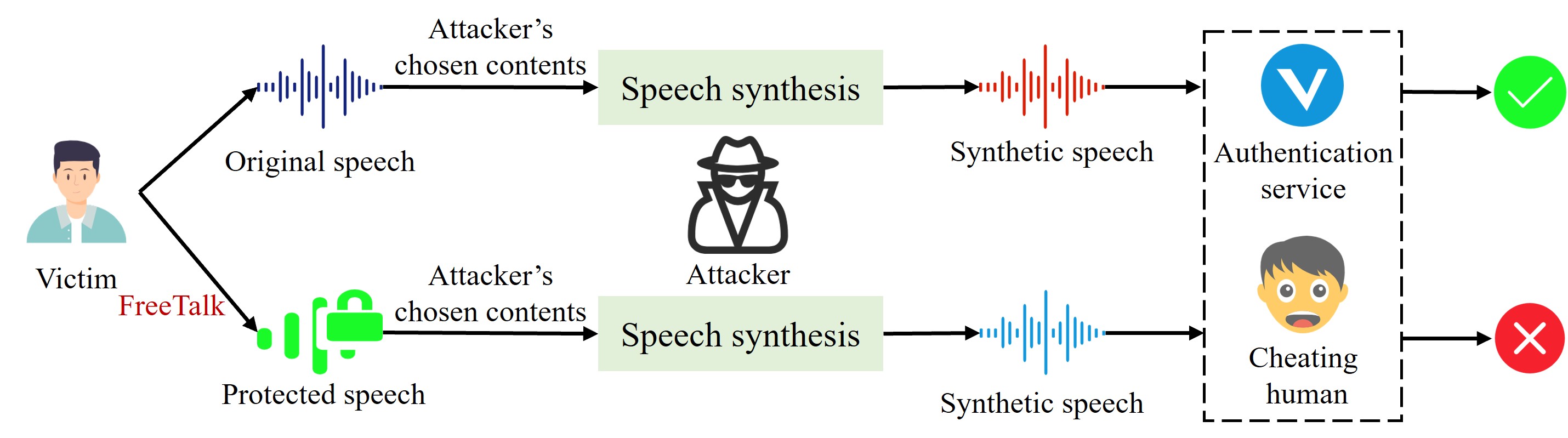}
\caption{Threat model of \name.}
\label{fig:TM}
\end{figure*}

\section{Design of \name}
In this section, we propose a sample-wise privacy-preserving method that can generate a specific perturbation for a piece of speech. Then, to ensure the robustness of the perturbation and speech quality, we also propose data augmentation strategy and noise smoothing mechanism to enhance the presented method. Furthermore, we present a novel identity-wise privacy protection method that can reduce computational overhead and support any length from the same speaker.
\begin{algorithm}[ht]
\caption{Sample-wise Protection}
\label{alg:sample-wise}
\begin{algorithmic}
\Require Source waveform $x$;  Frequency interval ratio $\alpha$;  Noise level $\lambda$;  Number of iterations $N$;
\Ensure Protected waveform $\hat{x}$

\Function{FrequencyBandProtection}{$x$}
    \State $e \gets \text{ExtractEmbedding}(x)$
    \State $s \gets \text{STFT}(x)$, $\overline{s} \gets |s|$
    \State // Define frequency band for perturbation
    \State $f_{s}, f_{e} \gets \text{ComputeFreqBand}(s,\alpha)$
    \State $T_{mask} \gets \min(\overline{s}[f_{start}:f_{end}])$
    
    \State // Initialize noise
    \State $\delta_{r}, \delta_{i} \gets \mathcal{N}(0,1)$
    \For{$step = 1$ to $N$}
        \State // Smooth and combine noise
        \State $\delta \gets \text{SmoothNoise}(\delta_{r} + j\cdot\delta_{i})$
        \State // Create and apply frequency mask
        \State $\text{mask}_{i, j}$ = $\begin{cases} 
(\overline{s} > T_{mask})_{i,j}, \quad\text{if}\quad j\in[f_{s}, f_{e}] \\ 0, \quad\text{otherwise} \end{cases}$
        \State $\hat{x} \gets \text{ISTFT}(s + \lambda \cdot \delta \cdot \text{mask})$
        \State $\hat{e} \gets \text{ExtractEmbedding}(\text{Augment}(\hat{x}))$
        \State // accumulate gradient
        \State Update $\delta$ with $\nabla_{\delta}\mathcal{L}(e,\hat{e})$ using Adam
    \EndFor
    
    \State \Return $\hat{x}$
\EndFunction
\end{algorithmic}
\end{algorithm}

\subsection{Sample-wise Optimization for Protection} Our proposed algorithm addresses unauthorized voice cloning by introducing frequency domain perturbations that preserve speech intelligibility while disrupting the extraction of speaker-specific features. This is motivated by the observation that TTS systems primarily rely on specific frequency bands to extract speaker characteristics~\cite{casanova2022yourtts, kim2020glow}, making these bands ideal targets for protective perturbation. Since speaker characteristics are predominantly encoded in specific frequency bands of the speech signal~\cite{Wang2017Tacotron}, here we strategically introduce controlled perturbations in these frequency ranges. In this way, we can effectively protect the speaker's identity while maintaining the natural quality of the speech. The core protection strategy involves the following key components.

Given an input waveform $x \in \mathbb{R}^{1\times L}$, we first extract its speaker embedding $e$ using a pre-trained speaker verification model. This embedding serves as a reference for measuring the effectiveness of our protection mechanism. Next, we compute its Short-Time Fourier Transform (STFT) representation $s \in \mathbb{C}^{F\times T}$ for the frequency feature, where $F$ and $T$ denote frequency bins and time frames, respectively. The detailed parameters of the STFT used in this paper are presented in ~\ref{Experiments and Evaluation}.A.6). The frequency band ($f_{s} = F\cdot\frac{1-\alpha}{2}$ and $f_{e} = F\cdot\frac{1+\alpha}{2}$) for perturbation is determined by a ratio parameter $\alpha$, which helps to identify the most relevant frequency components for the characteristics of the speaker. Within this band, we establish a masking threshold $T_{mask}$ based on the minimum magnitude of the STFT coefficients:
\begin{equation}
    T_{mask} = \min(|s|[f_{s}:f_{e}])
\end{equation}
which ensures that perturbations are applied only to significant spectral components. 

Further, we initialize the complex noise $\delta = \delta_r + j\cdot\delta_i$ where $\delta_r, \delta_i \sim \mathcal{N}(0,1)^{F \times T}$ represent the real and imaginary components, respectively, denoting the perturbation on the frequency domain. The noise is refined through $N$ iterations using gradient descent. At each iteration, we apply a frequency mask defined as:
\begin{equation}
\text{mask}_{i,j} = \begin{cases} (|s| > T_{mask})_{i,j}, & \text{if } j \in [f_s, f_e] \\
0, & \text{otherwise} \end{cases}
\end{equation}
This mask ensures that perturbations are concentrated in the target frequency band and applied only to significant spectral components. The protected waveform $\hat{x}$ is generated by applying the masked noise to the original spectrogram: $\hat{x} = \text{ISTFT}(s + \lambda \cdot \delta \cdot \text{mask})$, where $\lambda$ controls the perturbation magnitude. The effectiveness is evaluated by comparing the original embedding $e$ with the embedding $\hat{e}$ extracted from the augmented protected waveform (discussed in Sec.\ref{sec:augmentation}). The noise $\delta$ is updated using Adam optimizer to maximize the embedding distance while maintaining signal integrity using noise smoothing (discussed in Sec.\ref{sec:smoothing}) before applying the frequency perturbations. Our approach incorporates data augmentation and noise smoothing techniques to enhance robustness against various audio transformations. The augmentation (which might degrade audio quality) ensures protection persists under common audio processing operations while smoothing (which might sacrifice effectiveness) prevents artifacts and maintains natural sound quality. The iterative optimization process allows for fine-tuned control over the protection-intelligibility trade-off through parameters $\lambda$ and $N$, while the frequency-selective approach ensures minimal impact on the original speech content (the complete algorithm is presented in Algorithm.~\ref{alg:sample-wise}). This makes our method particularly suitable for real-world applications where both privacy protection and speech quality are crucial.

\subsection{Augmentation} \label{sec:augmentation}
The data augmentation strategy for waveform signals consists of three sequential transformations to enhance the robustness of protection. \textbf{Gaussian Noise Injection}: First, we add random gaussian noise to simulate environmental interference and prevent overfitting: $ x \sim \mathcal{N}(x, 0.01^2)$. This helps the perturbation become more resilient to background noise in real-world scenarios. \textbf{Time Shifting}: We then apply a random temporal shift to make the perturbation invariant to the exact timing of speech segments:
\begin{equation}
x_{1,i} = 
\begin{cases} 
x_{1,i-k}, & \text{if }0< k \leq i < L+k\\
0, & \text{if } i \geq L+k \text{ and } k > 0 \\
x_{1,i-k}, & \text{if } 0< -k \leq i < L-k \\
0, & \text{if } i < -k \text{ and } k < 0 \\
x_{1,i}, & \text{if } k = 0
\end{cases}
\end{equation}
where $k \sim \lfloor\mathcal{U}(-0.1\cdot L, 0.1\cdot L)\rfloor$. This transformation ensures that the model focuses on the speech content rather than its absolute position in time. \textbf{Volume Scaling}: Finally, we apply random amplitude scaling to account for varying speech volumes:
$ x = \kappa \cdot x$, where $\kappa \sim \mathcal{U}(0.8, 1.2)$.
This makes the model more robust to variations in speaking volume and recording conditions. The final augmented waveform incorporates all these transformations, creating a more challenging and diverse training signal. This comprehensive augmentation strategy helps prevent overfitting and improves the model's generalization ability across different acoustic conditions, speaking styles, and recording environments. The carefully chosen parameters (e.g., noise variance of 0.01, shift range of ±10\%, and scaling range of 0.8-1.2) provide meaningful perturbations while preserving the essential characteristics of the speech signal.

\subsection{Noise Smoothing}  \label{sec:smoothing}
The noise smoothing mechanism is designed to reduce high-frequency artifacts and ensure spectral consistency in the frequency domain. Given complex noise $\delta = \delta_r + j\cdot\delta_i$, we apply a one-dimensional averaging filter: (1) \textbf{Kernel Definition}. We first define a normalized averaging kernel: $\textbf{k} = \frac{1}{K}[1, 1, ..., 1] \in \mathbb{R}^K$, where we use $K = 5$ in this paper. This kernel performs local averaging over a window of 5 time steps. (2) \textbf{Smoothing Operation}. The smoothing is applied separately to real and imaginary components through valid convolution:
\begin{equation}
    \delta_r^{\text{s}} = (\delta_r \otimes \textbf{k})_{valid}, \delta_i^{\text{s}} = (\delta_i \otimes \textbf{k})_{valid}
\end{equation}
where $\otimes$ denotes the convolution operation with appropriate padding to maintain the original dimensions. Thus we have $\delta^{\text{s}} = \delta_r^{\text{s}} + j\cdot\delta_i^{\text{s}}$. This smoothing process serves multiple purposes: (1) Ensure temporal continuity in the perturbation signal; (2) Reduce potential artifacts that could arise from abrupt noise changes; (3) Maintain the spectral structure of the original signal while applying adversarial perturbations; (4) Help create more natural-sounding modifications by avoiding sharp transitions.

The smoothed noise is then applied to the original STFT representation with proper masking and scaling, resulting in more coherent and perceptually acceptable modifications to the audio signal. The kernel size of 5 provides a good balance between smoothing effectiveness and preservation of local temporal structure.

\subsection{Identity-wise Optimization for Protection} While sample-wise protection provides effective defense against voice cloning, it requires generating unique perturbations for each audio sample, which may be computationally intensive in real-world applications. To address this limitation, we propose an identity-wise protection mechanism that learns a universal perturbation pattern for each speaker identity, significantly reducing computational overhead while maintaining robust protection. Moreover, considering that the length of different waveforms varies as well, we introduce an identity-wise and frame-wise patch optimization strategy, enabling the preservation of waveforms with any length.

Our approach is motivated by the observation that speaker characteristics remain relatively consistent across different utterances from the same speaker. This consistency allows us to develop a single, optimized frequency perturbation that can effectively protect all audio samples from a given speaker. Given a training set $X = \{x_1, ..., x_n\}$ containing multiple utterances from the same speaker, we learn a universal complex noise pattern $\delta = \delta_r + j\cdot\delta_i$ that generalizes across all samples. The algorithm consists of two main components.

\textbf{Universal Noise Learning.} 
Similarly, we initialize identity-wise complex noise components $\delta_r, \delta_i \sim \mathcal{N}(0,1)^{F\times T}$ and optimize them iteratively. For each training iteration, instead of applying direct gradient descent, we employ a PGN~\cite{2023PGN}-inspired two-stage interpolation strategy \textbf{to better what}. Specifically, for each audio sample $x_i$, we first sample a perturbed noise pair $(\delta_r', \delta_i')$ from a local neighborhood and compute the first-stage gradient $\nabla_{\delta}^{(1)} = \nabla_{\delta} \mathcal{L}(e_i, \hat{e}_i')$. Then, we perform a forward projection to generate $\delta^\star = \delta' - \alpha \cdot \nabla_{\delta}^{(1)}$, and obtain the second-stage gradient $\nabla_{\delta}^{(2)} = \nabla_{\delta} \mathcal{L}(e_i, \hat{e}_i^\star)$ with the updated protection. The final gradient is interpolated as:
\begin{equation}
\nabla_{\delta} = (1 - \delta) \cdot \nabla_{\delta}^{(1)} + \delta \cdot \nabla_{\delta}^{(2)},
\end{equation}
which is accumulated across all training samples. This interpolation strategy stabilizes the optimization and encourages the learned universal noise to be more transferable and robust.

\textbf{Frame-wise Noise Application.} To enhance robustness and reduce artifacts, we introduce a novel frame-wise noise application strategy. Specifically, we reformulate the perturbation into the tiled frame patch, which is applicable for waveforms of any length. Given frame length $l$ (patch length) and mask ratio $r$, we first divide the STFT spectrogram $s$ into $P = \lfloor |s|/l \rfloor$ frames. Then, a random binary mask is generated where each frame has probability $(1-r)$ of being selected:

\begin{equation}
\text{mask}_i = 
\begin{cases}
1, & \text{if } \text{Random}(0,1) \geq r \\
0, & \text{otherwise}
\end{cases}
\end{equation}
With the mask, we apply the universal noise selectively to chosen frames: $s_{m:n} = s_{m:n} + \lambda \cdot \delta \cdot \text{mask}_i$, where $m = (i-1) \cdot l$ and $n = i \cdot l$ define frame boundaries.

Several key parameters control the effectiveness of our method: the noise level $\lambda$ affecting protection strength, the mask ratio $r$ controlling the density of protection, and the frame length of patch $l$ determining the granularity of noise application. These parameters can be tuned to achieve an optimal balance between protection effectiveness and speech quality for different application scenarios, which will be further discussed in our ablation study. And to further smooth high-frequency artifacts, we apply a lightweight Wiener filter to the perturbed STFT, enhancing perceptual quality without degrading protection.

This frame-wise application strategy serves multiple purposes: (1) introduces temporal variability in protection, making it harder for attackers to reverse-engineer the protection; (2) reduces potential artifacts by maintaining clean frames; (3) allows for fine-grained control over the protection-quality trade-off through the mask ratio $r$.

The overall process for identity-wise optimization is shown in Algorithm.\ref{alg:identity-wise}. This approach offers several advantages over sample-wise protection:
\begin{itemize}
    \item Reduced computational cost during deployment, as the noise pattern needs to be computed only once per speaker.
    \item Consistent protection across all utterances with various lengths from the same speaker.
\end{itemize}


    

    

\begin{algorithm}
\caption{Identity-wise Protection}
\label{alg:identity-wise}
\begin{algorithmic}
\Require Train Audio set $X = \{x_1, ..., x_n\}$; Frame length $l$; Noise level $\lambda$; Mask ratio $r$; Number of iterations $N$
\Ensure Universal noise $\delta_{r}, \delta_{i}$

\Function{UniversalFrequencyProtection}{$X$}
    \State // Initialize complex universal noise
    \State $\delta_{r}, \delta_{i} \gets \mathcal{N}(0,1)^{F \times l}$
    \For{$step = 1$ to $N$}
        \State $\nabla \gets 0$
        \For{$x_i \in X$}
            \State $e_i \gets \text{ExtractEmbedding}(x_i)$
            \State $\nabla^{(i)} \gets 0$
            \For{$k = 1$ to $K$}
                \State Sample: $\delta_{r}', \delta_{i}' \gets \delta_{r}, \delta_{i} + \mathcal{N}(0, \epsilon)$
                \State $\hat{x}_i' \gets \text{AddFreqNoise}(x_i, \delta_{r}', \delta_{i}', r)$
                \State $\hat{e}_i' \gets \text{ExtractEmbedding}(\hat{x}_i')$
                \State $\nabla^{(1)} \gets \nabla_{\delta} \mathcal{L}(e_i, \hat{e}_i')$
                \State Update $(\delta_r^\star, \delta_i^\star)$ using Adam with $\nabla^{(1)}$
                \State $\hat{x}_i^\star \gets \text{AddFreqNoise}(x_i, \delta_r^\star, \delta_i^\star, r)$
                \State $\hat{e}_i^\star \gets \text{ExtractEmbedding}(\hat{x}_i^\star)$
                \State $\nabla^{(2)} \gets \nabla_{\delta} \mathcal{L}(e_i, \hat{e}_i^\star)$
                \State $\nabla^{(i)} \gets \nabla^{(i)} + (1-\delta) \cdot \nabla^{(1)} + \delta \cdot \nabla^{(2)}$
            \EndFor
            \State $\nabla \gets \nabla + \frac{1}{K} \cdot \nabla^{(i)}$
        \EndFor
        \State Update $(\delta_{r}, \delta_{i})$ using Adam with $\nabla$
    \EndFor
    \State \Return $\delta_{r}, \delta_{i}$
\EndFunction

\Function{AddFreqNoise}{$x, \delta_{r}, \delta_{i}, r$}
    \State $s \gets \text{STFT}(x)$, $\delta \gets \delta_{r} + j \cdot \delta_{i}$
    \State $P \gets \lfloor |s|/l \rfloor$, $\text{mask} \gets \text{RandMask}(r, P)$
    \If{all(mask$=0$)} 
        \State Randomly set one $\text{mask}[j] \gets 1$
    \EndIf
    \For{$i = 1$ to $P$}
        \If{$\text{mask}[i] = 1$}
            \State $m \gets (i-1) \cdot l$, $n \gets i \cdot l$
            \State $s_{m:n} \gets s_{m:n} + \lambda \cdot \delta$
        \EndIf
    \EndFor
    \State $s \gets \text{WienerFilter}(s)$
    \State \Return $\text{ISTFT}(s)$
\EndFunction
\end{algorithmic}
\end{algorithm}

\section{Experiments and Evaluation}
\label{Experiments and Evaluation}
\subsection{Experimental Settings}
All experiments were conducted on a server equipped with Intel(R) Xeon(R) Platinum 8358P CPUs, 3.40GHz processor, 386GB RAM, and NVIDIA A800. VSCode and PyTorch are used to deploy the model and complete relevant experiments.
\subsubsection{Speech Corpus Datasets}
Because of the ethical considerations, we use public datasets (LibriSpeech and VoxCeleb1) to evaluate the effectiveness of \name.

\textbf{LibriSpeech}\cite{panayotov2015librispeech} is a corpus of read English speech which contains $1000$ hours of speech sampled at $16$kHz. It is suitable for training and evaluating speech recognition systems. LibriSpeech is frequently used to train and evaluate automatic speech recognition (ASR) models. Due to its large scale and accurate text annotations, it has become one of the standard datasets in ASR research.

\textbf{VoxCeleb1}\cite{nagrani2017voxceleb} is an audio-visual dataset extracted from interview videos uploaded to YouTube, which contains more than hundreds of thousands of 'real world' utterances for over $1000$ celebrities. Due to its diversity of speakers and background noise, VoxCeleb1 is particularly valuable for developing robust speaker recognition systems. It is often used to benchmark the performance of speaker identification and verification models.

\subsubsection{Speech Synthesis Models}
In our threat model, we assume that the defender has no knowledge about the speech synthesis models used by the attacker. Hence, to verify the defense performance of \name, we also employ five state-of-the-art speech synthesis models, including Tacotron2-DDC\cite{Tacotron2-DDC}, Tacotron2-DCA\cite{gorodetskii2022zero}, Glow-TTS\cite{kim2020glow}, YourTTS\cite{casanova2022yourtts}, and Speedy-Speech\cite{vainer2020speedyspeech}. 

\textbf{Tacotron2-DDC}\cite{Tacotron2-DDC} is proposed based on TacoTron2\cite{shen2018natural} that synthesizes time-domain waveforms by using a modified WaveNet vocoder. Tacotron2-DDC applys Dynamic Decode Convolution enhancement into decoder, which enables the model to capture both local and global contextual information more effectively, leading to improved speech synthesis quality.

\textbf{Tacotron2-DCA}\cite{gorodetskii2022zero} is a popular zero-shot speech synthesis model, which achieves better text alignment and smoother, natural speech synthesis by using an energy-based attention mechanism. This model can synthesize high-quality speech for victim speakers only with a few seconds of target audio.

\textbf{Glow-TTS}\cite{kim2020glow} is stream-based generation model specifically designed for parallel Text-To-Speech (TTS) conversion. It supports end-to-end alignment and faster inference to produce stable and consistent output.

\textbf{YourTTS}\cite{casanova2022yourtts} is a zero-shot multi-speaker TTS and voice conversion model, which is suitable for high-quality cross-language speech synthesis. It can generate the synthetic audio with less than one minute voice of the victim.

\textbf{Speedy-Speech}\cite{vainer2020speedyspeech} is a lightweight TTS model with fast inference speed, which is suitable for the real-time applications. It  can be efficiently trained on a single GPU and can run in real time even on a CPU.

\subsubsection{Speech Verification Models}
In our paper, we suppose that both speech verification (SV) model and speech synthesis model have a similar function (feature extraction) as an encoder. Therefore, we use a speech verification model (e.g., X-Vector\cite{parcollet2022speechbrain,speechbrain}) to optimize the protected victim's audio. Moreover, to verify whether a synthesis voice based on the protected audio matches the victim's original audio, we also employ four pre-trained speech verification models (ECAPA\cite{parcollet2022speechbrain,speechbrain}, ResNet\cite{parcollet2022speechbrain,speechbrain}, ERes2Net\cite{chen20243d, 3D-Speaker}, CAM++\cite{chen20243d, 3D-Speaker}) to evaluate. 

\textbf{X-Vector}\cite{parcollet2022speechbrain,speechbrain} is a classic DNN-based speaker verification model by using convolutional layers and average pooling for fixed-length embedding. Data augmentation (e.g., adding noises and reverberation) is also proposed to improve the model performance and robustness.

\textbf{ECAPA}\cite{parcollet2022speechbrain,speechbrain} is a Time Delay Neural Network (TDNN)-based speaker feature extractor, which puts more emphasis on channel attention, propagation and aggregation for high-precision feature extraction.

\textbf{ResNet}\cite{parcollet2022speechbrain,speechbrain} can capture rich speaker features, particularly robust in complex audio environments because of the residual network structure.

\textbf{ERes2Net}\cite{chen20243d, 3D-Speaker} is the extended Res2Net with multi-scale feature learning, capturing speaker info effectively even in short segments.

\textbf{CAM++}\cite{chen20243d, 3D-Speaker} is a speaker recognition model based on densely connected delay neural network, which uses CAM-Pooling Plus for adaptive feature aggregation, enhancing verification performance. 

\subsubsection{Speech Recognition Model}
To evaluate the impact of \name on speech recognition task, we use Whisper \cite{whisper} to convert the original speech and protected speech to text, respectively. Whisper is a multilingual speech recognition model developed by OpenAI. Whisper is an end-to-end model with multilingual and multitasking capabilities, including speech-to-text, speech translation, and speaker recognition.

\subsubsection{Evaluation Metrics}
In order to evaluate the privacy-preserving performance of our method, we use Source and TTS-Camouflage Waveforms Mismatch Rate (STCMR) and Source and TTS-Camouflage Waveforms Similarity (STCS) to show the difference between the source audio and the synthetic voice. Moreover, to evaluate the speech quality of the protected audio, we also employ Mean Option Score (MOS). Besides, to evaluate the utility of the protected audio, we also use Connectionist Temporal Classification Loss (CTC-LOSS), Character Error Rate (CER) and Word Error Rate (WER). The introduction of the evaluation metrics is shown as following.

\textbf{STCMR} is defined as the rate that the generated synthetic audio based on the protected voice can not be matched with the source audio of the original speaker after inputting the SV model, which likes the false recognition rate of the SV model. The value of STCMR is $0$ to $1$. A higher STCMR value represents a better privacy-preserving effectiveness, which indicates that the more difference between the source audio and the synthetic audio.

\textbf{STCS} is defined as the similarity between the synthetic audio and the source audio after feature extraction, whose scale is from $0$ to $1$. A low STCS denotes that the more difference between the source audio and the synthetic audio, which also shows a better defense performance.

\textbf{MOS} represents a ranking of the quality and naturalness of speech on a scale from $0$ to $5$. This metric is used to assess the overall audio quality of the protected audio. The higher MOS indicates the better quality of the generated audio.



\textbf{CTC-LOSS} is the Connectionist Temporal Classification Loss. We take the original audio as the input and the Whisper output as the real label. CTC-LOSS represents the difference between the protected audio after inputting the Whisper and the real label in the feature space. When the CTC-LOSS is less than 10, it indicates that the prediction result is similar to the real label. That is, the lower CTC-LOSS represents better quality speech.

\textbf{CER} is the Character Error Rate that represents the character difference between the original audio and the protected audio after being converted to text. The lower CER represents better quality speech.

\textbf{WER} is the Word Error Rate that represents the word difference between the original audio and the protected audio after being converted to text. The lower WER represents better quality speech.

Therefore, to ensure defense performance, speech quality, and audio utility simultaneously, our goal is to achieve higher STCMR, MOS, and lower STCS, CTC-LOSS, CER, WER.
\subsubsection{Parameters Configuration}
In our experiments, we configure our method to optimize privacy-preserving speaker embeddings through a series of adversarial steps and data augmentations. The primary parameters for our experimental setup are as follows:

\textbf{Details of STFT}: We adopt Short-Time Fourier Transform (STFT) for frequency-domain processing of waveform audio. In all experiments, the following parameter settings are used:
\begin{itemize}
    \item \textbf{Sampling rate:} 16,000\,Hz
    \item \textbf{FFT window size:} 1024
    \item \textbf{Hop length:} 512
    \item \textbf{Window length:} 1024
    \item \textbf{Window function:}  Hann window\footnote{The Hann window is a commonly used tapering function that reduces spectral leakage by smoothly weighting the ends of each frame towards zero.}
\end{itemize}

The STFT is computed on single-channel audio. The same configuration is used for both STFT and inverse STFT operations. These settings ensure consistent time-frequency resolution and phase preservation throughout all frequency-domain manipulations.

\textbf{Loss Function}: We used cosine similarity as loss function, an approach that helps ensure that the resulting embedding differs from the original without needing a specific target. This untargeted setup supports general privacy protection without narrowing the embedding to match a particular output.
    
\textbf{Optimization Steps}: The model undergoes $100$ optimization steps to iteratively refine the waveform embeddings. Each step applies small perturbations to the waveform in the frequency domain.
    
\textbf{Learning Rate}: The learning rate is set to $0.1$ for the Adam optimizer to ensure controlled yet effective convergence during the refinement process.

\textbf{Frequency Interval}: We set the frequency interval to $0.6$, enabling perturbations focused around the middle frequency range. This choice limits the noise impact on the signal while maximizing privacy in critical frequency bands.

\textbf{Data Augmentation}: Data augmentation is applied to enhance the robustness of the refined waveform. This augmentation includes minor noise additions, time shifting, and random scaling to simulate varied acoustic environments.

\textbf{Noise Smoothing}: A smoothing operation is applied to the perturbation noise, with a kernel size of $5$, to reduce abrupt changes in the frequency domain. This noise smoothing enhances the perceptual quality.
    
\textbf{Noise Level}: We set the noise level to $0.05$ for each perturbation applied to the frequency spectrum. This controlled noise level balances embedding robustness and signal quality.

\subsection{Overall Performance for \name}

In our study, we employ the X-Vector to generate the perturbation. Four speech synthesis models and four speech verification models are used to evaluate the privacy-preserving performance of the perturbation generated by our method. The overall performance of the proposed approach is shown in Table \ref{tab:overall_performance}. From the experimental results in Table \ref{tab:overall_performance}, we mainly have three observations. 

\textbf{(1) Privacy-preserving performance}: we observe that the lowest STCMR exceeds $0.96$ and the highest STCS is not more than $0.15$ for the four speech synthesis models and four speech verification models on VoxCeleb1. It is difficult for speech verification models to recognize the synthetic audio and the original audio as the voice of the same person. This indicates that the synthetic audio generated based on the protected speech differs greatly from the original audio in feature space, demonstrating the effectiveness of our method. Moreover, we find that the protective performance of our method is somewhat different in different datasets. It is obvious that the protection performance is better on VoxCeleb1 rather than LibriSpeech. For example, the lowest STCMR values are $0.96$ and $0.67$, respectively. We suspect the reason is that VoxCeleb1 contains more natural speech data from conversations, interviews, speeches, and other scenarios, making the speech content closer to everyday life and actual communication contexts, possibly including various background noises, filler words, pauses, etc., with richer and more diverse emotional expressions and language styles. This makes the optimized perturbations with better transferability. In contrast, the speech content of LibriSpeech mainly consists of readings of written language, with relatively standard and formal language expression, and a relatively single emotional tone, resulting in poorer transferability. Furthermore, the performance of \name can be seen in different voice authentication models. In addition, we observe that the protection performance has subtle differences in various speech verification models. For example, when facing YourTTS on the LibriSpeech, the STCMR is 0.71 and 0.91 on the ECAPA and ERes2Net,respectively. We guess that it is due to ERes2Net placing more emphasis on the fusion extraction of local and global features, integrating features within a single residual block to extract local signals, and aggregating different scale acoustic features from various hierarchical outputs as global signals, thereby obtaining more representative and distinctive speaker characteristics. In contrast, ECAPA utilizes a delay neural network framework to extract features at different time steps to capture the temporal information of speech, and then further filters and enhances key features through mechanisms such as SESqueeze Excitation Layer, making the speaker's feature information less susceptible to disturbance. Finally, we can find that \name has a stable defense performance on different speech synthesis models. For example, the STCMR's fluctuation range under different speech synthesis models is no more than $0.1$ when operating on VoxCeleb1 and CAM++. 

\textbf{(2) Speech quality}: For audio protected by our method, we can see that the MOS is from $3.15$ to $3.23$. It objectively shows the quality of the audio processed by our method. 

\textbf{(3) Speech utility}: To demonstrate the utility of protected speech, we input the protected speech and the original speech into Whisper, respectively. From Table \ref{tab:overall_performance}, we can know that CTC-LOSS remains at a high value, indicating that the protected speech and the original speech are similar in the feature space after Whisper processing. Additionally, when converting the original speech and the protected speech into text using Whisper, it was found that both the character error rate and the word error rate are 0. This shows that protected speech has almost no impact on the speech conversion task and does not significantly compromise speech usability.

\begin{table*}[!htp]
    \centering
    \caption{Overall performance of \name.}
    \scalebox{0.78}{
    \renewcommand{\arraystretch}{1.5}
    \begin{tabular}{cccccccccccccc}
         \hline
         \multirow{2}{*}{Dataset} & \multirow{2}{*}{TTS Model} & \multicolumn{2}{c}{ECAPA}  & \multicolumn{2}{c}{ResNet} & \multicolumn{2}{c}{ERes2Net} & \multicolumn{2}{c}{CAM++} & \multirow{2}{*}{MOS} & \multirow{2}{*}{CTC-LOSS} & \multirow{2}{*}{CER} & \multirow{2}{*}{WER}\\ 
         \cline{3-10}
         &  & STCMR & STCS & STCMR & STCS & STCMR & STCS & STCMR & STCS &  &  &  &\\
         \hline
         \multirow{5}{*}{VoxCeleb1} & Glow-TTS & 0.97±0.00 & 0.08±0.01 & 0.99±0.00 & 0.09±0.01 & 1.00±0.00 & 0.11±0.01 & 1.00±0.00 & 0.14±0.01 & 3.21±0.23 & \multirow{5}{*}{3.65±1.37} & \multirow{5}{*}{0.00±0.00} & \multirow{5}{*}{0.00±0.00} \\
         & Speedy-Speech & 0.96±0.00 & 0.09±0.01 & 0.99±0.00 & 0.10±0.01 & 0.99±0.00 & 0.12±0.01 & 0.99±0.00 & 0.15±0.01 & 3.16±0.24 & & &\\
         & Tacotron2-DCA & 0.97±0.00 & 0.08±0.01 & 0.99±0.00 & 0.10±0.01 & 1.00±0.00 & 0.12±0.01 & 1.00±0.00 & 0.14±0.01 & 3.20±0.25 & & &\\
         & Tacotron2-DDC & 0.96±0.00 & 0.10±0.01 & 0.98±0.00 & 0.11±0.01 & 0.99±0.00 & 0.13±0.02 & 0.99±0.00 & 0.15±0.01 & 3.15±0.20 & & &\\
         & YourTTS & 0.97±0.00 & 0.09±0.01 & 0.99±0.00 & 0.10±0.01 & 1.00±0.00 & 0.13±0.02 & 1.00±0.00 & 0.14±0.02 & 3.19±0.22 & & &\\
         \hline
         \multirow{5}{*}{LibriSpeech} & Glow-TTS & 0.72±0.00 & 0.23±0.01 & 0.84±0.00 & 0.20±0.01 & 0.95±0.00 & 0.22±0.01 & 0.81±0.00 & 0.27±0.01 & 3.21±0.20 &\multirow{5}{*}{4.15±0.91} & \multirow{5}{*}{0.00±0.00} & \multirow{5}{*}{0.00±0.00}\\
         & Speedy-Speech & 0.73±0.00 & 0.23±0.01 & 0.85±0.00 & 0.20±0.01 & 0.95±0.00 & 0.22±0.01 & 0.82±0.00 & 0.26±0.01 & 3.19±0.19 & & &\\
         & Tacotron2-DCA & 0.67±0.00 & 0.24±0.01 & 0.81±0.00 & 0.22±0.01 & 0.91±0.00 & 0.24±0.01 & 0.74±0.00 & 0.29±0.01 & 3.17±0.20 & & &\\
         & Tacotron2-DDC & 0.71±0.00 & 0.24±0.01 & 0.84±0.00 & 0.21±0.01 & 0.91±0.00 & 0.24±0.01 & 0.79±0.00 & 0.28±0.01 & 3.17±0.19 & & &\\
         & YourTTS & 0.71±0.00 & 0.22±0.01 & 0.84±0.00 & 0.20±0.01 & 0.91±0.00 & 0.24±0.01 & 0.80±0.00 & 0.26±0.01 & 3.23±0.22 & & &\\
         \hline
    \end{tabular}}
    \label{tab:overall_performance}
\end{table*}
\subsection{Sensitivity Analysis}
In this subsection, we evaluate various of factors that may influence the privacy-preserving performance of our method. 

\subsubsection{Different noise levels}
\begin{figure*}[!htp]
    \centering
    \subcaptionbox{Privacy Metric.\label{fig:noise_privacy}}[0.48\textwidth]{
        \includegraphics[width=\linewidth]{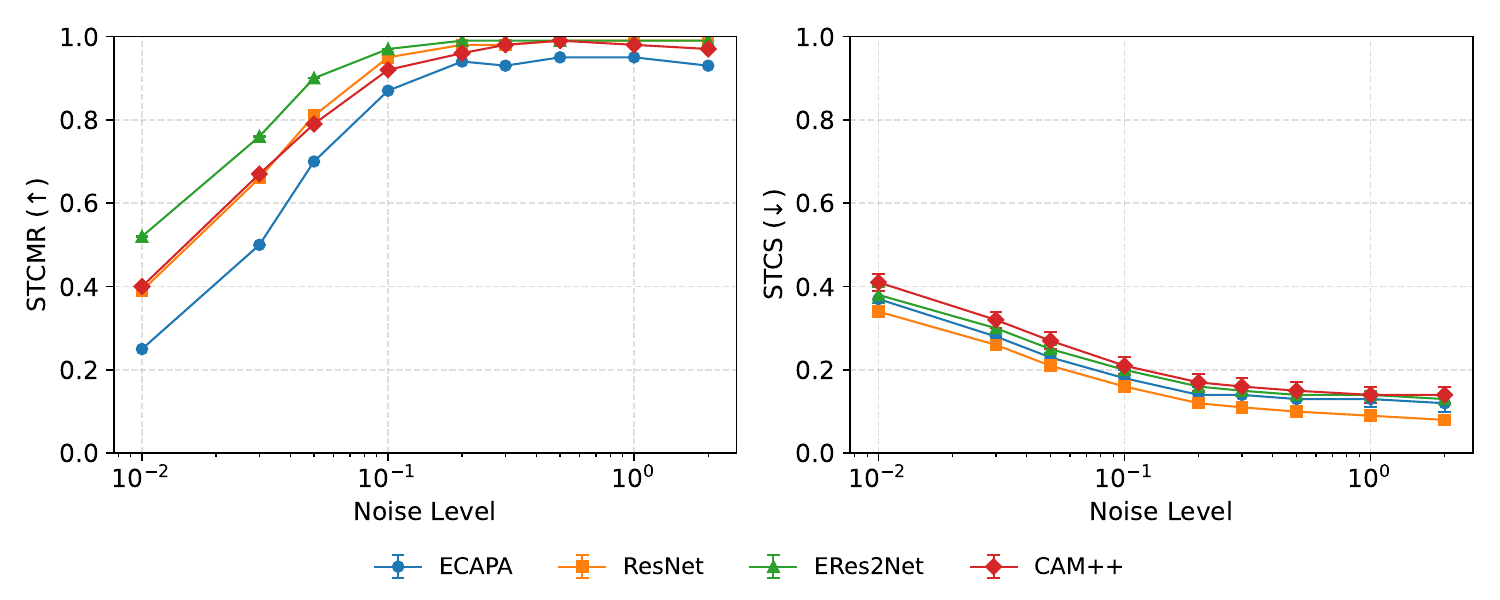}
    }
    \hfill
    \subcaptionbox{Quality Metric.\label{fig:noise_quality}}[0.48\textwidth]{
        \includegraphics[width=\linewidth]{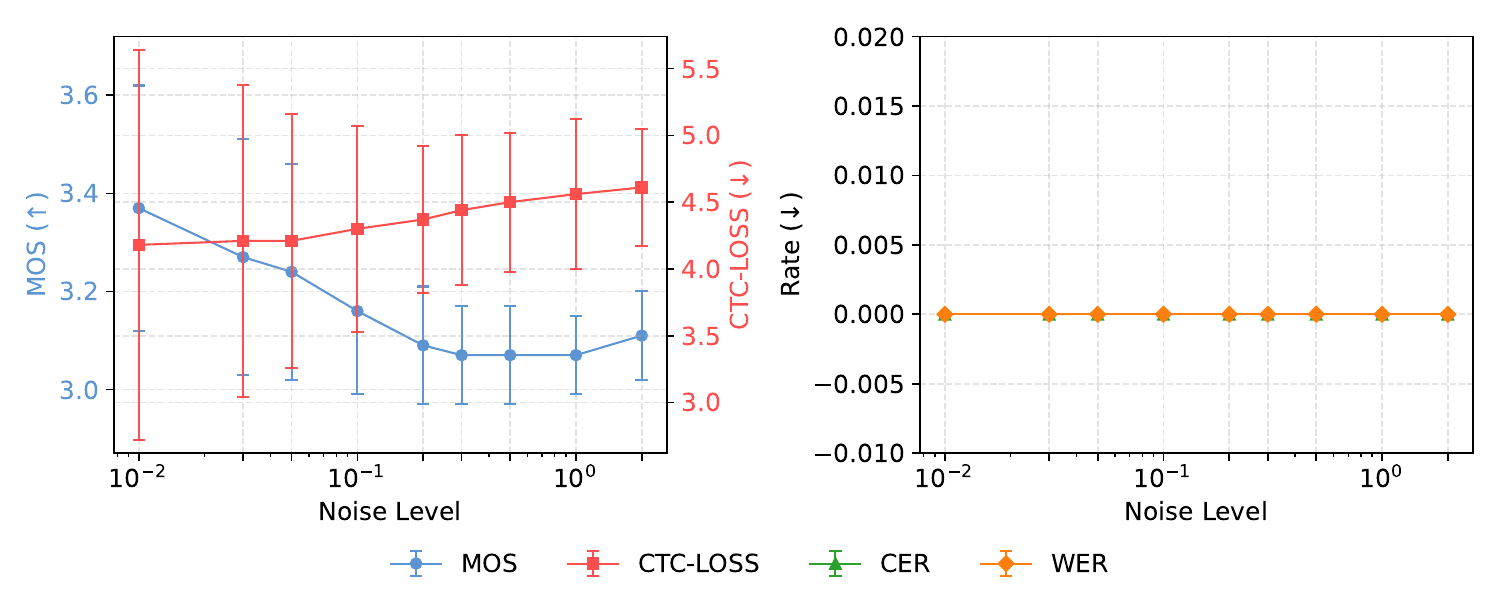}
    }

    \caption{Privacy-preserving performance of \name against YourTTS with different noise levels on LibriSpeech.}
    \label{tab:different_noise_levels}
\end{figure*}

Our method has to generate the the the perturbation added to the frequency spectrum. The noise level of the perturbation may affect the performance of our method. Therefore, we generate perturbation based on the X-Vector and evaluate the defense performance with different noise levels when facing YourTTS on LibriSpeech. As shown in Fig.~\ref{tab:different_noise_levels}, with an increase in noise level, the value of STCMR and STCS maintains a general growth and decreases, respectively. It indicates that protection performance keeps improving as the noise level increases. However, the MOS is gradual decline, which indicates that the speech quality is deterioration. We infer that as the noise level increases, the noise has a greater effect on the audio. While the privacy-preserving is improved, the sound quality is also compromised. Moreover, for the speech utility, we can see that both CER and WER always maintain 0. It indicates that the different noise levels nearly have no influence on the speech utility.

\begin{figure*}[!htp]
    \centering
    \subcaptionbox{Privacy Metric.\label{fig:freq_privacy}}[0.48\textwidth]{
        \includegraphics[width=\linewidth]{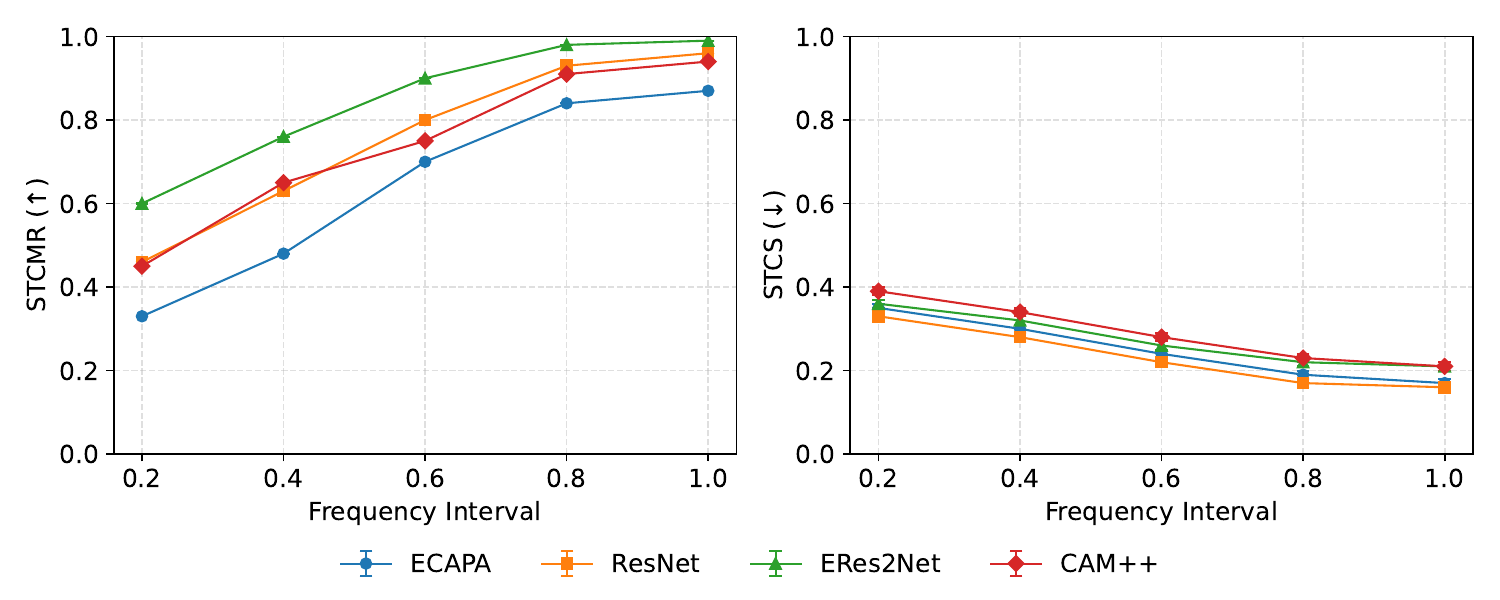}
    }
    \hfill
    \subcaptionbox{Quality Metric.\label{fig:freq_quality}}[0.48\textwidth]{
        \includegraphics[width=\linewidth]{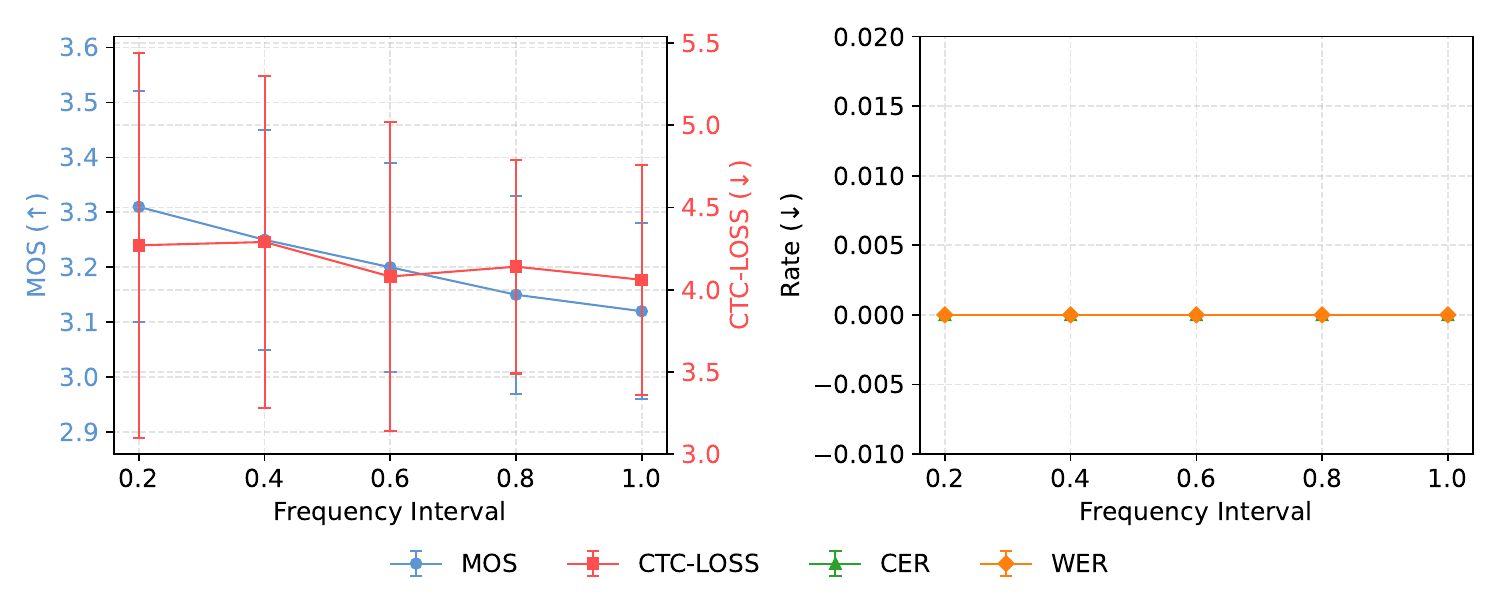}
    }

    \caption{Privacy-preserving performance of \name on different frequency intervals on YourTTS and LibriSpeech.}
    \label{tab:different_frequency_intervals}
\end{figure*}

\subsubsection{Different frequency intervals}
Considering that our method needs to add a perturbation to the frequency domain of the source waveform, the value of the frequency interval may influence the performance of our method. Therefore, keeping the dataset, the speech synthesis model and the speech verification model consistent with those in the previous paragraph (different noise level), we evaluate the performance of our methods in different frequency intervals when the noise level is fixed at $0.05$. The experimental results are shown in Fig.~\ref{tab:different_frequency_intervals}. From Fig.~\ref{tab:different_frequency_intervals}, we can see that with an increase in the frequency interval value, the STCMR and STCS continue to rise and fall, respectively. It indicates that the privacy-preserving effect is improving steadily. Moreover, the MOS drops gradually, which demonstrates that the audio quality gradually deteriorates. We guess that with the increasing of frequency interval, the window for adding noise becomes larger. For the YourTTS model, the features become more blurred, and this also may lead to a decrease in sound quality. Besides, CTC-LOSS is almost stable. Both CER and WER always maintain 0. It indicates that the different frequency intervals have almost no influence on the speech utility.

\subsubsection{Different speech verification models}
In our method, the defender needs to use a speech verification model to assist in generating the perturbation. To verify the generalization of our method, we evaluated the privacy-preserving performance on four different models and VoxCeleb1 to optimize the perturbation. Other parameter settings are set the same as the overall experiments. The experimental results are shown in Table \ref{tab:different_speech_verification_models}. From Table \ref{tab:different_speech_verification_models}, we can know when the perturbation is generated based on different models, the STCMR and STCS have some fluctuations. For example, when facing Glow-TTS, the protected audio generated by ECAPA (STCMR:0.91±0.00, STCS:0.18±0.01) can achieve a more privacy-preserving performance than that generated by ERes2Net  (STCMR:0.72±0.00, STCS:0.23±0.01). However, the lowest STCMR is still more than 0.72 and the highest STCS is less than 0.32. This indicates that our method has a good generalization on different models. In addition, for speech quality, we can see that the MOS is almost stable around 3.2. This demonstrates that the speech quality is almost unaffected by speech verification model. For speech utility, the CTC-LOSS is always around 3.8. Moreover, both CER and WER maintain 0. This indicates that the speech utility is not affected by different speech verification models in our method.

\begin{table*}[!htp]
    \centering
    \caption{Privacy-preserving performance of \name on different speech verification models on VoxCeleb1.}
    \scalebox{0.75}{
    \renewcommand{\arraystretch}{1.5}
    \begin{tabular}{cccccccccccccc}
         \hline
        \multirow{2}{*}{SVM} & \multirow{2}{*}{TTS Model} & \multicolumn{2}{c}{ECAPA}  & \multicolumn{2}{c}{ResNet} & \multicolumn{2}{c}{ERes2Net} & \multicolumn{2}{c}{CAM++} & \multirow{2}{*}{MOS} & \multirow{2}{*}{CTC-LOSS} & \multirow{2}{*}{CER} & \multirow{2}{*}{WER} \\ 
         \cline{3-10}
         &  & STCMR & STCS & STCMR & STCS & STCMR & STCS & STCMR & STCS &  &  & & \\
         \hline
         \multirow{5}{*}{CAM++} & Glow-TTS & 0.80±0.00 & 0.23±0.01 & 0.87±0.00 & 0.24±0.01 & 0.92±0.00 & 0.28±0.01 & 0.90±0.00 & 0.31±0.01 & 3.20±0.24 & \multirow{5}{*}{3.78±1.67} &  \multirow{5}{*}{0.00±0.00} &  \multirow{5}{*}{0.00±0.00}\\
         & Speedy-Speech & 0.78±0.00 & 0.22±0.01 & 0.86±0.00 & 0.23±0.01 & 0.88±0.00 & 0.27±0.01 & 0.88±0.00 & 0.30±0.01 & 3.26±0.24 & & &\\
         & Tacotron2-DCA & 0.83±0.00 & 0.22±0.01 & 0.87±0.00 & 0.24±0.01 & 0.90±0.00 & 0.28±0.01 & 0.89±0.00 & 0.31±0.01 & 3.31±0.28 & & &\\
         & Tacotron2-DDC & 0.82±0.00 & 0.22±0.01 & 0.86±0.00 & 0.25±0.01 & 0.91±0.00 & 0.28±0.01 & 0.89±0.00 & 0.31±0.01 & 3.27±0.23 & & &\\
         & YourTTS & 0.84±0.00 & 0.20±0.01 & 0.89±0.00 & 0.22±0.01 & 0.92±0.00 & 0.28±0.01 & 0.92±0.00 & 0.30±0.01 & 3.33±0.33 & & &\\
         \hline
         
         \multirow{5}{*}{ECAPA} & Glow-TTS & 0.91±0.00 & 0.18±0.01 & 0.93±0.00 & 0.18±0.01 & 0.97±0.00 & 0.22±0.01 & 0.97±0.00 & 0.26±0.01 & 3.17±0.27 & \multirow{5}{*}{3.82±2.54} & \multirow{5}{*}{0.00±0.00} & \multirow{5}{*}{0.00±0.00}\\
         & Speedy-Speech & 0.90±0.00 & 0.18±0.01 & 0.94±0.00 & 0.18±0.01 & 0.97±0.00 & 0.22±0.01 & 0.96±0.00 & 0.25±0.01 & 3.15±0.21 & & &\\
         & Tacotron2-DCA & 0.90±0.00 & 0.19±0.01 & 0.93±0.00 & 0.19±0.01 & 0.96±0.00 & 0.23±0.01 & 0.98±0.00 & 0.26±0.01 & 3.24±0.25 & & &\\
         & Tacotron2-DDC & 0.89±0.00 & 0.19±0.01 & 0.92±0.00 & 0.19±0.01 & 0.97±0.00 & 0.22±0.01 & 0.95±0.00 & 0.25±0.01 & 3.16±0.19 & & &\\
         & YourTTS & 0.91±0.00 & 0.17±0.01 & 0.94±0.00 & 0.18±0.01 & 0.96±0.00 & 0.22±0.01 & 0.96±0.00 & 0.25±0.01 & 3.23±0.28 & & &\\
         \hline
         
         \multirow{5}{*}{ERes2Net} & Glow-TTS & 0.72±0.00 & 0.23±0.01 & 0.76±0.00 & 0.25±0.01 & 0.83±0.00 & 0.29±0.01 & 0.83±0.00 & 0.32±0.01 & 3.30±0.32 & \multirow{5}{*}{3.89±2.47} & \multirow{5}{*}{0.00±0.00} & \multirow{5}{*}{0.00±0.00}\\
         & Speedy-Speech & 0.72±0.00 & 0.23±0.01 & 0.77±0.00 & 0.24±0.01 & 0.85±0.00 & 0.29±0.01 & 0.84±0.00 & 0.31±0.01 & 3.23±0.26 & & &\\
         & Tacotron2-DCA & 0.72±0.00 & 0.22±0.01 & 0.78±0.00 & 0.25±0.01 & 0.86±0.00 & 0.29±0.01 & 0.85±0.00 & 0.31±0.01 & 3.31±0.32 & & &\\
         & Tacotron2-DDC & 0.75±0.00 & 0.22±0.01 & 0.79±0.00 & 0.25±0.01 & 0.87±0.00 & 0.29±0.01 & 0.85±0.00 & 0.31±0.01 & 3.28±0.27 & & &\\
         & YourTTS & 0.80±0.00 & 0.21±0.01 & 0.84±0.00 & 0.23±0.01 & 0.87±0.00 & 0.28±0.01 & 0.86±0.00 & 0.30±0.01 & 3.37±0.37 & & &\\
         \hline

         \multirow{5}{*}{ResNet} & Glow-TTS & 0.81±0.00 & 0.20±0.01 & 0.86±0.00 & 0.22±0.01 & 0.91±0.00 & 0.26±0.01 & 0.92±0.00 & 0.29±0.01 & 3.20±0.29 & \multirow{5}{*}{3.74±1.26} & \multirow{5}{*}{0.00±0.00} & \multirow{5}{*}{0.00±0.00}\\
         & Speedy-Speech & 0.82±0.00 & 0.20±0.01 & 0.88±0.00 & 0.22±0.01 & 0.92±0.00 & 0.26±0.01 & 0.91±0.00 & 0.28±0.01 & 3.27±0.32 & & &\\
         & Tacotron2-DCA & 0.83±0.00 & 0.20±0.01 & 0.86±0.00 & 0.23±0.01 & 0.92±0.00 & 0.27±0.01 & 0.94±0.00 & 0.29±0.01 & 3.32±0.32 & & &\\
         & Tacotron2-DDC & 0.81±0.00 & 0.21±0.01 & 0.87±0.00 & 0.23±0.01 & 0.92±0.00 & 0.27±0.01 & 0.93±0.00 & 0.30±0.01 & 3.32±0.30 & & &\\
         & YourTTS & 0.85±0.00 & 0.20±0.01 & 0.86±0.00 & 0.22±0.01 & 0.96±0.00 & 0.27±0.01 & 0.94±0.00 & 0.29±0.01 & 3.28±0.31 & & &\\
         \hline
         
    \end{tabular}}
    \label{tab:different_speech_verification_models}
\end{table*}

\subsection{Ablation Study}
In our method, we employ data augmentation technology to enhance the model robustness and use the noise smoothing mechanism to improve the speech quality. Therefore, we analyze the impact of data augmentation and noise smoothing on our method. We generate a perturbation by using the X-Vector on the LibriSpeech. YourTTS is used to evaluate the effectiveness of our method. As shown in Table \ref{tab:data_augmentation_and_noise_smoothing}, we can see that when only employing data augmentation technology, the STCMR increases and the STCS decreases. Meanwhile, the MOS also slightly declines. It indicates that using data augmentation can improve the privacy-preserving performance while reducing the audio quality in our method. Moreover, when only using the noise smoothing mechanism, the STCMR decreases, while the STCS increases. This demonstrates that the privacy-preserving effect is reduced while the audio quality improving when the noise smoothing mechanism is used. Hence, to balance the privacy-preserving performance and audio quality, both data augmentation technology and the noise smoothing mechanism are used in our method. 

\begin{table*}[!htp]
    \centering
    \caption{Privacy-preserving performance of the \name on LibriSpeech and YourTTS for whether using data augmentation and noise smoothing.}
    \scalebox{0.75}{
    \renewcommand{\arraystretch}{1.5}
    \begin{tabular}{cccccccccccccc}
         \hline
        \multirow{2}{*}{Data Augmentation} & \multirow{2}{*}{Noise Smoothing} & \multicolumn{2}{c}{ECAPA}  & \multicolumn{2}{c}{ResNet} & \multicolumn{2}{c}{ERes2Net} & \multicolumn{2}{c}{CAM++} & \multirow{2}{*}{MOS} & \multirow{2}{*}{CTC-LOSS} & \multirow{2}{*}{CER} & \multirow{2}{*}{WER}  \\ 
         \cline{3-10}
         & & STCMR & STCS & STCMR & STCS & STCMR & STCS & STCMR & STCS &  &  &  &\\
         \hline
         FALSE & FALSE & 0.84±0.00 & 0.17±0.01 & 0.97±0.00 & 0.15±0.01 & 0.97±0.00 & 0.19±0.01 & 0.91±0.00 & 0.20±0.01 & 3.37±0.21 & 4.38±0.58 & 0.00±0.00 & 0.00±0.00\\
         TRUE & FALSE & 0.90±0.00 & 0.16±0.01 & 0.98±0.00 & 0.14±0.01 & 0.99±0.00 & 0.18±0.01 & 0.94±0.00 & 0.19±0.01 & 3.30±0.21 & 4.30±0.59 & 0.00±0.00 & 0.00±0.00\\
         FALSE & TRUE & 0.63±0.00 & 0.25±0.01 & 0.81±0.00 & 0.22±0.01 & 0.87±0.00 & 0.26±0.01 & 0.78±0.00 & 0.28±0.01 & 3.30±0.22 & 4.21±1.14 & 0.00±0.00 & 0.00±0.00\\
         TRUE & TRUE & 0.70±0.00 & 0.23±0.01 & 0.82±0.00 & 0.21±0.01 & 0.91±0.00 & 0.25±0.01 & 0.79±0.00 & 0.27±0.01 & 3.23±0.22 & 4.15±0.91 & 0.00±0.00 & 0.00±0.00\\
         \hline
    \end{tabular}}
    \label{tab:data_augmentation_and_noise_smoothing}
\end{table*}
\subsection{Transferability Analysis}
The existing methods have to generate a specific perturbation for each audio sample. In this paper, to reduce the computational overhead of the user, we proposed an identity-wise optimization protection method. In this section, we evaluate the performance of the proposed identity-wise optimization protection method. Moreover, we also consider two situations: the frame length is 30 and 120, respectively. Concretely, we use ECAPA to assist in generating the protected audio and employ YourTTS to evaluate the defense performance on LibriSpeech. We divided the corpus of one person into a training set and a test set. Then, we generate a universal perturbation patch based on the training dataset. Finally, we add the patch to the train set and the test set, respectively. Relevant indicators are also used to evaluate the effectiveness of \name. This section considers the noise size ranges from 0.1 to 0.5 and the noise frame length is 30 and 120, respectively. The relevant experimental results are shown in Table \ref{tab:transferability analysis}. Three conclusions can be drawn as follows:

\textbf{(1) Privacy-preserving}: Note that Train-STCMR and Test-STCMR represent the STCMR on the training set and test set, respectively. Train-STCS and Test-STCS represent the STCS on the training set and test set, respectively. From Table \ref{tab:transferability analysis}, when the noise frame length is 30, we can know that both Train-STCMR and Test-STCMR maintain around $0.90$. This indicates that the synthetic audio and the original audio are identified as different people. Moreover, Train-STCS and Test-STCS are always stable from $0.13$ to $0.27$, indicating that the synthetic audio and the original audio have great differences in feature space. Compared with the experimental results with noise frame length=30, when the noise frame length is 120, we can see that both Train-STCMR and Test-STCMR slightly decrease, Train-STCS and Test-STCS slightly increase. It indicates that privacy-preserving performance is better when noise frame length is 30. In addition, note that when the noise frame length is 120 and the noise level is 0.1 and 0.2, both Train-STCMR and Test-STCMR are significantly lower than the other results. We guess that when the noise frame length is 120, many samples were truncated. When the noise level is 0.1 and 0.2, respectively, the noise amplitude is relatively small. This has a significant impact on the effect of privacy protection. Overall, most of Train-STCMR and Test-STCMR are above 0.8. most of Train-STCS and Test-STCS are lower than 0.3. Therefore, it indicates that \name can achieve satisfactory privacy-preserving performance.

\textbf{(2) Speech quality and utility}: After adding the universal perturbation patch, protected audio maintains high speech quality and utility. From Table \ref{tab:transferability analysis}, we can know that most of MOS remain above 2.5. It should be noted that when the noise frame length is 120, the overall MOS is higher than that when the noise frame length is 30. We indicate that under the same noise level conditions, when the noise frame length is longer, the added noise amplitude is lower. This results in better audio quality and a higher MOS value. Moreover, we can see that the CER and WER always remain at 0. This indicates that the noise frame length has little impact on the speech utility. The experimental results indicate that the universal perturbation patch generated by our method has some influence on speech quality and negligible influence on speech utility.

\textbf{(3) Others}: The proposed identity-wise optimization protection method is not sensitive to noise level and the speech verification model. From Table \ref{tab:transferability analysis}, we can see that the fluctuations of STCMR and STCS on the training set and the test set are small with the gradual increase of noise level. This indicates that the privacy protection performance is stable. Moreover, the MOS shows a little decreasing trend, indicating that the protected speech quality gradually decreased with the increase of noise level. However, the CER and WER are almost stable, indicating that there is little impact on the utility of the protected audio. Besides, the experimental results remain stable on the different speech verification models, which indicates that our method has great robustness.

\begin{table*}[!htp]
    \centering
    \caption{Privacy-preserving performance of the proposed identity-wise optimization protection method on different speech verification models with LibriSpeech and YourTTS.}
    \resizebox{1.8\columnwidth}{!}{
    \scalebox{0.82}{
    \renewcommand{\arraystretch}{1.5}
    \begin{tabular}{ccccccc|ccccc}
         \hline
         \multirow{2}{*}{SVM} & \multirow{2}{*}{Metric} & \multicolumn{5}{c|}{Noise Frame Length=30} & \multicolumn{5}{c}{Noise Frame Length=120} \\ 
         \cline{3-12}
         & & 0.1 & 0.2 & 0.3 & 0.4 & 0.5 & 0.1 & 0.2 & 0.3 & 0.4 & 0.5\\
         \hline
         \multirow{4}{*}{ECAPA} & Train-STCMR & 0.84±0.00 & 0.72±0.00 & 1.00±0.00 & 0.77±0.00 & 1.00±0.00 & 0.27±0.00 & 0.62±0.00 & 0.83±0.00 & 0.93±0.00 & 0.78±0.00 \\
         & Train-STCS & 0.20±0.01 & 0.25±0.01 & 0.13±0.00 & 0.22±0.01 & 0.16±0.01 & 0.34±0.01& 0.26±0.01 & 0.21±0.01 & 0.16±0.01 & 0.21±0.00 \\
         & Test-STCMR & 0.86±0.00 & 0.72±0.00 & 1.00±0.00 & 0.86±0.00 & 1.00±0.00 & 0.13±0.00& 0.72±0.00 & 0.82±0.00 & 0.96±0.00 & 0.86±0.00 \\
         & Test-STCS & 0.23±0.01 & 0.26±0.01 & 0.14±0.00 & 0.22±0.00 & 0.14±0.01 &  0.38±0.01 & 0.27±0.01 & 0.23±0.01 & 0.17±0.00 & 0.21±0.00 \\
         \hline

        \multirow{4}{*}{ResNet} & Train-STCMR & 0.86±0.00 & 0.80±0.00 & 1.00±0.00 & 0.93±0.00 & 1.00±0.00 & 0.31±0.00 & 0.65±0.00 & 0.84±0.00 & 0.93±0.00 & 0.92±0.00 \\
         & Train-STCS & 0.21±0.01 & 0.23±0.01 & 0.13±0.00 & 0.23±0.00 & 0.16±0.01 & 0.35±0.01 & 0.28±0.01 & 0.24±0.01 & 0.17±0.01 & 0.21±0.01 \\
         & Test-STCMR & 0.89±0.00 & 0.89±0.00 & 1.00±0.00 & 0.89±0.00 & 1.00±0.00 & 0.27±0.00 & 0.62±0.00 & 0.82±0.00 & 1.00±0.00 & 0.93±0.00 \\
         & Test-STCS & 0.23±0.01 & 0.24±0.01 & 0.14±0.01 & 0.22±0.01 & 0.13±0.00 & 0.37±0.01 & 0.29±0.01 & 0.23±0.01 & 0.17±0.01 & 0.22±0.01 \\
         \hline
        
        \multirow{4}{*}{ERes2Net} & Train-STCMR & 0.96±0.00 & 0.93±0.00 & 1.00±0.00 & 0.96±0.00 & 1.00±0.00 & 0.60±0.00 & 0.89±0.00 & 0.96±0.00 & 0.96±0.00 & 0.95±0.00 \\
         & Train-STCS & 0.22±0.01 & 0.25±0.00 & 0.12±0.01 & 0.23±0.01 & 0.18±0.01 & 0.36±0.01 & 0.29±0.01 & 0.26±0.01 & 0.19±0.01 & 0.25±0.01 \\
         & Test-STCMR & 0.89±0.00 & 0.96±0.00 & 1.00±0.00 & 1.00±0.00 & 1.00±0.00 & 0.48±0.00 & 0.93±0.00 & 0.96±0.00 & 1.00±0.00 & 1.00±0.00\\
         & Test-STCS & 0.24±0.01 & 0.27±0.01 & 0.12±0.00 & 0.23±0.00 & 0.17±0.00 & 0.39±0.01 & 0.29±0.01 & 0.27±0.01 & 0.22±0.00 & 0.24±0.01\\
         \hline

         \multirow{4}{*}{CAM++} & Train-STCMR & 0.95±0.00 & 0.87±0.00 & 1.00±0.00 & 0.93±0.00 & 1.00±0.00 & 0.62±0.00 & 0.77±0.00 & 0.90±0.00 & 0.95±0.00 & 0.92±0.00\\
         & Train-STCS & 0.25±0.00 & 0.26±0.01 & 0.16±0.00 & 0.25±0.01 & 0.20±0.01 & 0.35±0.01 & 0.31±0.01 & 0.28±0.01 & 0.24±0.01 & 0.26±0.01 \\
         & Test-STCMR & 0.93±0.00 & 0.96±0.00 & 1.00±0.00 & 0.96±0.00 & 1.00±0.00 & 0.34±0.00 & 0.86±0.00 & 0.93±0.00 & 0.96±0.00 & 0.86±0.00 \\
         & Test-STCS & 0.27±0.00 & 0.27±0.01 & 0.18±0.00 & 0.26±0.00 & 0.19±0.00 & 0.40±0.01 & 0.31±0.01 & 0.28±0.01 & 0.25±0.01 & 0.27±0.01\\
         \hline
         
         \multicolumn{2}{c}{MOS} & 2.74±0.03 & 2.53±0.02 & 2.25±0.02 & 2.50±0.01 & 2.17±0.01 & 3.00±0.09 & 2.84±0.05 & 2.87±0.03 & 2.62±0.01 & 2.81±0.02\\
         \multicolumn{2}{c}{CER} & 0.00±0.00 & 0.00±0.00 & 0.00±0.00 & 0.00±0.00 & 0.00±0.00 & 0.00±0.00 & 0.00±0.00 & 0.00±0.00 & 0.00±0.00 & 0.00±0.00 \\
         \multicolumn{2}{c}{WER} & 0.00±0.00 & 0.00±0.00 & 0.00±0.00 & 0.00±0.00 & 0.00±0.00 & 0.00±0.00 & 0.00±0.00 & 0.00±0.00 & 0.00±0.00 & 0.00±0.00 \\
         \hline
    \end{tabular}}}
    \label{tab:transferability analysis}
\end{table*}

\begin{figure}[]
\centering
\includegraphics[width=0.8\linewidth]{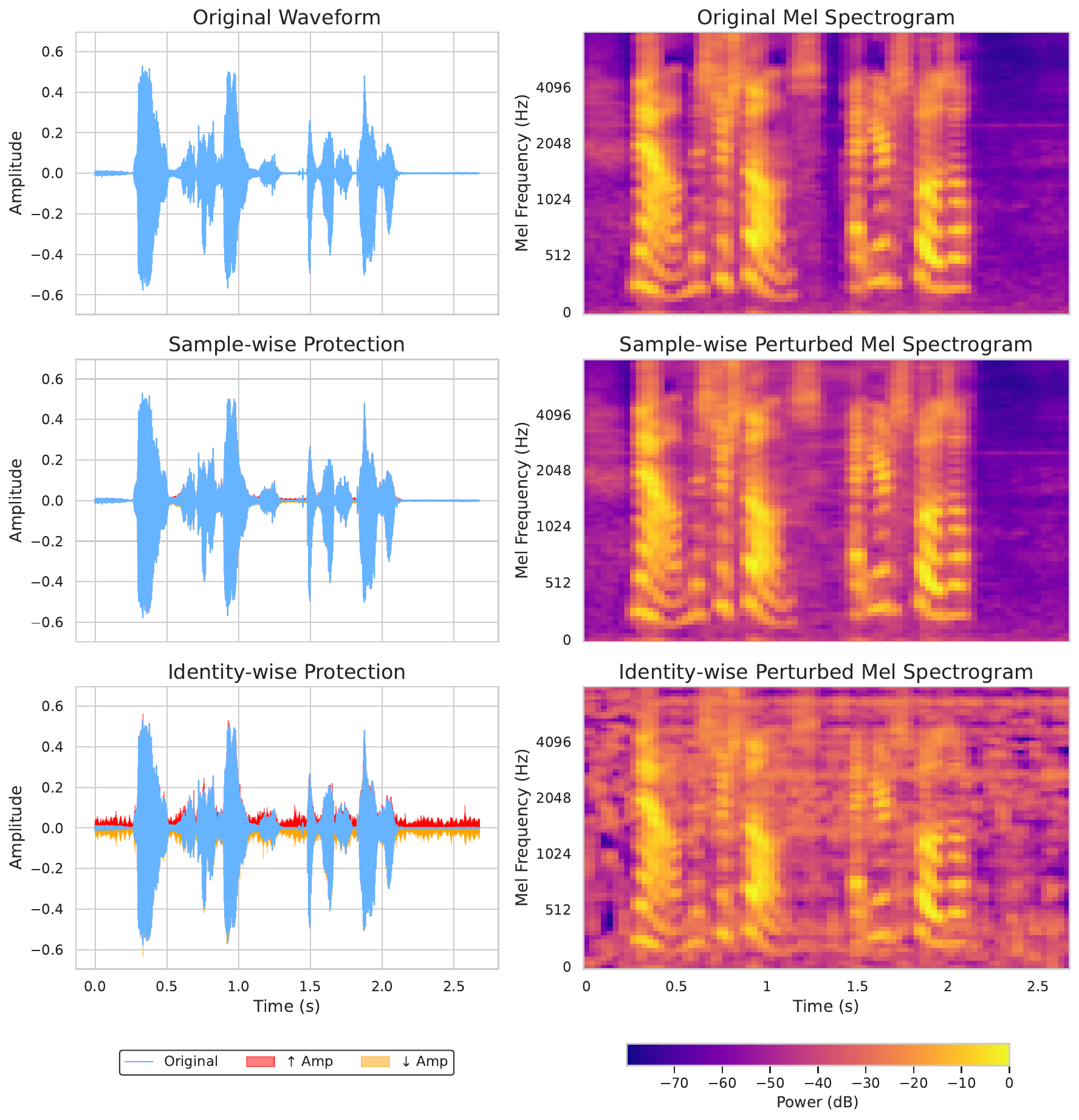}
\caption{Temporal and frequential domain comparison between original and \name-protected audios.}
\label{fig:distortion_visualization}
\end{figure}

\subsection{Distortion Analysis}
To evaluate the perceptual distortion of \name, we conduct a visual inspection of the protected audio in both temporal and frequential domains, as shown in Fig.~\ref{fig:distortion_visualization}. 

In the temporal plots, we observe that the waveforms of \name-protected audios remain closely aligned with the original signal. The overall phase and temporal progression are unaffected, with only subtle amplitude perturbations being introduced. These perturbations manifest as slight upward or downward shifts in local peaks and troughs, yet the global waveform envelope is preserved. Such characteristics indicate that the prosodic rhythm, syllable timing, and energy contours of speech are maintained, which are critical for intelligibility and the perception of natural speech. From a perceptual standpoint, this ensures that listeners do not experience noticeable artifacts or discontinuities. In the frequential domain, the Mel Spectrograms reveal distinct behaviors for the two protection strategies. For sample-wise protection, the perturbations appear as scattered, localized intensity changes across both low-frequency and mid-frequency bands. These fine-grained variations differ across individual utterances, reflecting the stochastic and instance-dependent nature of this protection. Importantly, these modifications do not obscure the broader formant structures or harmonic patterns, thereby preserving phonetic clarity while injecting sufficient randomness to hinder model inversion or feature reconstruction. By contrast, identity-wise protection introduces perturbations that are more consistent across the entire spectro-temporal span. Rather than appearing as isolated fluctuations, these perturbations form smoother, frame-spanning adjustments that shift the energy distribution of the spectrum in a controlled manner. This uniformity encodes protection signals that are stable across all utterances from the same speaker, effectively obfuscating speaker identity. While the deviations are slightly stronger than in the sample-wise case, they remain constrained in magnitude such that the essential timbre and resonance characteristics of the original voice are preserved. Consequently, the speech continues to sound natural to human listeners, but automatic speaker recognition models face greater difficulty in associating the protected audio with the true identity.

Overall, both protection modes strike a balance between robustness and imperceptibility: sample-wise perturbations emphasize subtle, transient distortions, while identity-wise perturbations embed stable, identity-level shifts. Together, these demonstrate that \name achieves effective protection with low perceptual degradation.


\section{Conclusion}
In this paper, we firstly propose a sample-wise privacy protection method against speech synthesis attacks. Our method protects user's voice characteristic by adding a frequency-domain perturbation. Then, to improve the transferability and robustness of our method, we also propose data augmentation strategy and noise smoothing mechanism to maintain natural speech quality. Finally, we also propose a novel identity-wise protection mechanism, which can support any length speech and generate a universal perturbation for a speaker. It should be noted that our method can perform well in a black-box setting. Extensive experimental results show the superior performance of \name on different models and datasets. The protected speech has satisfied privacy preservation performance, high speech quality and utility. In addition, the experimental results also demonstrate that the proposed \name has a fine generalization ability on different noise levels, different frequency intervals, and different speech verification models. The relevant experimental results also indicate that our method has better potential practicality for real-world scenarios.

\bibliographystyle{IEEEtran}
\bibliography{references}





\end{document}